\documentclass[pra,twocolumn,showpacs,aps,floatfix]{revtex4-1}
\usepackage{times}
\usepackage{amsmath}
\usepackage{amssymb}
\usepackage{amsfonts}
\usepackage{bm}
\usepackage{graphicx} 
\usepackage{epstopdf}
\usepackage{multirow}
\usepackage{hyperref} 
\newcommand{\ASBS}{New Zealand Foundation for Research, Science and Technology contracts NERF-UOOX0703, and UOOX0801}
 \newcommand{\eref}[1]{(\ref{#1})} 
\newcommand{\eeref}[1]{equation (\ref{#1})} 
\newcommand{\fref}[1]{Fig.~\ref{#1}}

\newcommand{\sref}[1]{section \ref{#1}}

\newcommand{\mbf}[1]{\mathbf{#1}}

\newcommand{\x}{\mathbf{x}}

\newcommand{\xp}{\mathbf{x}^\prime}

\newcommand{\p}{\mathbf{p}}

\newcommand{\eqn}[1]{\begin{eqnarray}#1 \end{eqnarray}}

\newcommand{\ecut}{\epsilon_{\rm{cut}}}

\newcommand{\ncut}{n_{\rm{cut}}}

\newcommand{\EQ}[1]{\begin{eqnarray}#1\end{eqnarray}}

\newcommand{\myint}[1]{\int d^3 #1\;}

\newcommand {\etal}{{\em et al.}}
\newcommand{\of}[1]{(#1)}
\newcommand{\rC}{\textbf{C}} 
\newcommand{\rI}{\textbf{I}} 

\newcommand{\PC}{\mathcal{P}_{\rC}}

\newcommand{\intV}[1]{\int d^3 #1\;}

\def\x{\mathbf{x}}

\begin{document}

\title{Decay of a quantum vortex: test of non-equilibrium theories for warm Bose-Einstein condensates}
\author{S. J. Rooney} 
\author{A.~S. Bradley} 
\email{corresponding author: abradley@physics.otago.ac.nz}
\author{P. B. Blakie}

\affiliation{Jack Dodd Center for Quantum Technology, Department of Physics, University of Otago, Dunedin, New Zealand.}
\date{\today}
\begin{abstract}
The decay of a vortex from a non-rotating high temperature Bose-Einstein condensate (BEC) is modeled using the stochastic projected Gross-Pitaevskii equation (SPGPE). In order to exploit the tunability of temperature in SPGPE theory while maintaining the total atom number constant, we develop a simple and accurate Hartree-Fock method to estimate the SPGPE parameters for systems close to thermal equilibrium. We then calculate the lifetime of a vortex using three classical field theories that describe vortex decay in different levels of approximation. The SPGPE theory is shown to give the most complete description of the decay process, predicting significantly shorter vortex lifetimes than the alternative theories. Using the SPGPE theory to simulate vortex decay for a trapped gas of $5\times 10^5$ $^{87}$Rb atoms, we calculate a vortex lifetime $\bar{t}$ that decreases linearly with temperature, falling in the range 20{\rm s}$ >\bar{t} >$1.5{\rm s} corresponding to the temperature range $0.78T_c\leq T\leq0.93T_c$. The vortex lifetimes calculated provide a lower bound for the lifetime of a persistent current with unit winding number in our chosen trap geometry, in the limit of vanishing vortex pinning potential.
\end{abstract}
\maketitle
Bose-Einstein condensates present a unique opportunity for developing \textit{ab initio} theory that can be directly compared to experiments. However, in the area of non-equilibrium dynamics most quantitative comparisons have been limited to the zero temperature regime, where the Gross-Pitaevskii equation (GPE) provides a comprehensive treatment.

An area of particular interest occurs in the non-equilibrium dynamics leading to the nucleation of vortices into a condensate, either as it is stirred or as it grows out of a rotating vapor, and the subsequent formation of a vortex lattice \cite{Madison2000,*Haljan2001,*Raman2001}. Two types of theoretical model have been used to  describe this process: (i) Gross-Pitaevskii treatments in which a damping parameter is introduced \cite{Penckwitt2002,Tsubota2002}, (ii) Hamiltonian classical-field methods, in which the nonlinearity within the Gross-Pitaevskii equation and appropriately chosen random initial conditions allows a stirred condensate to relax to a vortex lattice \cite{Lobo2004,Parker05a,Wright08a}.
While this theoretical work (note also the recent works using other approaches~\cite{Bradley08a,Dagnino09a}) has established qualitative understanding of the processes that lead to vortex lattice formation, the timescales predicted for lattice formation are qualitatively incomparable with experiments.  A general theory of vortex formation is further complicated by the variety of nucleation scenarios (a number of distinct mechanisms have recently been observed \cite{Hadzibabic2006a,*Sadler2006a,*Schweikhard2007a,*Scherer2007,*Ryu07a,*Lagoudakis08a,*Lin09a,Weiler08a}), and the challenges in developing a fully time dependent non-equilibrium theory of the interacting Bose gas that can account for all of the relevant physics at finite temperature. Indeed, the frontiers of many body physics~\cite{Bloch2008a}, and the application of generalized classical field theories~\cite{Blakie08a} to non-equilibrium~\cite{Scott08a,*Bezett09a,*Wright09b}, finite temperature~\cite{Wouters09a,*Bisset09a,*Bisset09c}, and critical phenomena~\cite{Bezett09b,*Bisset09b} in systems of ultra-cold Bosons are fields of much current interest.

In this paper we examine a well-defined process suitable for testing finite temperature theories: the decay of a singly quantized vortex in a stationary Bose-Einstein condensate at finite temperature.
From general arguments it is clear that such vortices are metastable in the absence of external rotation~\cite{Rokhsar97a,Fetter2001}, and will decay in the presence of dissipation. Previous work on this problem has included: a variational treatment based on the formalism of Stoof and coworkers \cite{Duine2004}, a pure classical-field treatment \cite{Schmidt2003}, simulations using the Zaremba-Nikuni-Griffin formalism \cite{Jackson09a}, and using a damped GPE \cite{Madarassy08a}. We also refer to Ref.~\cite{Wild09a} for a self consistent meanfield treatment, which has been limited to low temperatures, with the prediction that  the vortex precesses but does not decay.

\begin{figure}[!tb]
\begin{center}
\includegraphics[width=\columnwidth]{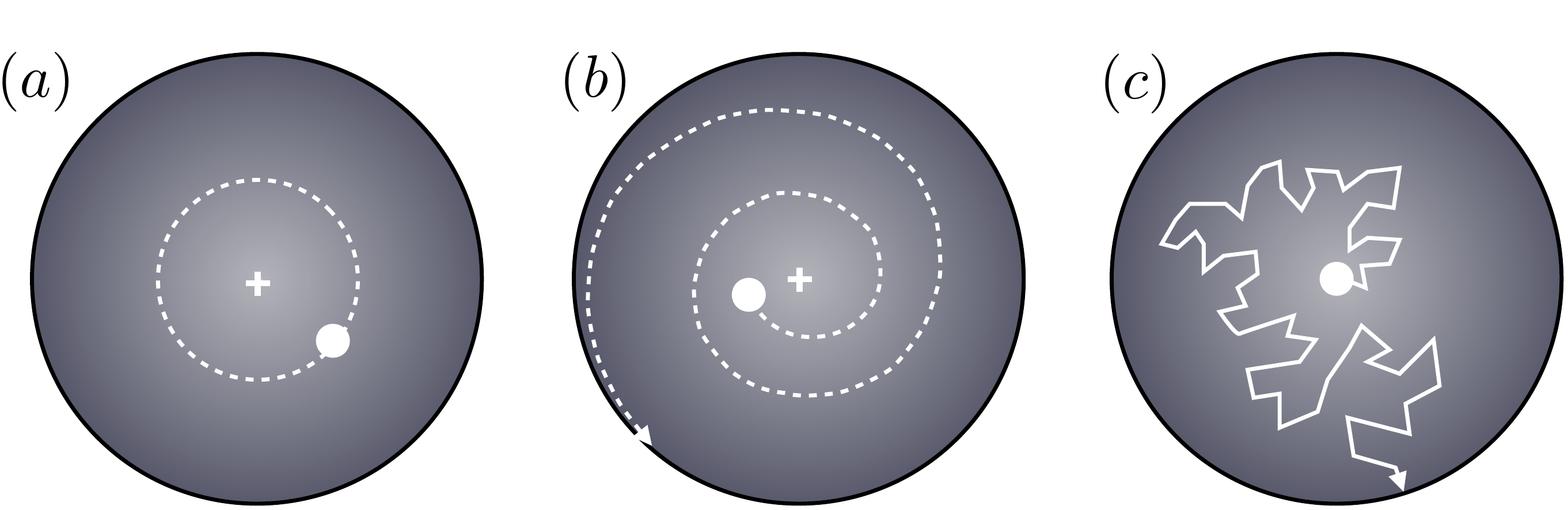}
\caption{Schematic of different vortex decay processes in two dimensions.  (a) Under Hamiltonian evolution, an eigenstate of the GPE containing a displaced vortex will evolve so that the vortex follows trajectories of constant radius about the centre of the condensate. At finite temperatures, the presence of a non-rotating thermal cloud creates a damping mechanism for vortex motion. (b) \emph{Dissipative} decay: in the absence of noise, a damped, off center vortex will decay to the edge of the condensate along a regular spiral trajectory. (c) \emph{Diffusive} decay: thermal fluctuations lead to stochastic vortex dynamics resembling Brownian motion. }
\label{fig1}
\end{center}
\end{figure}

 In the present work we use the stochastic projected Gross-Pitaevskii equation (SPGPE) \cite{SGPEI,SGPEII,Bradley08a} to study vortex decay. We show, by neglecting various processes, that this formalism can be reduced to the damped GPE and classical field treatments \cite{Schmidt2003,Madarassy08a}. In more detail, our approach is a c-field theory (see \cite{Blakie08a}) in which the modes of the system are divided into two regions according to occupation: appreciably occupied (low energy) modes constitute the coherent region and are described by a classical field theory based on the Wigner phase space representation; the remaining sparsely occupied modes constitute the incoherent region which behaves like a thermalized reservoir and is treated semi-classically. In the full SPGPE treatment the dynamics of the coherent region are simulated according to a stochastic differential equation that includes damping and noise arising from the incoherent region~\cite{Bradley08a}.  By neglecting this noise a damped projected Gross-Pitaevskii equation (DPGPE) treatment is obtained (similar to that in \cite{Madarassy08a,Jackson09a}). By neglecting both the noise and damping from the incoherent region (i.e.~all coupling to the incoherent region) the projected Gross-Pitaevskii equation (PGPE) classical field treatment is obtained (similar to that in \cite{Schmidt2003}).
 
By comparing simulations of the SPGPE, DPGPE, and PGPE, we are able to understand the relationships between these theories and
to investigate various physical processes that contribute to vortex  decay. 
In the absence of any dissipative/thermal effects the vortex can only precess in the system and it is unable to leave (Fig.~\ref{fig1}(a)). The damping effect of thermal cloud leads to a dissipative decay of the vortex (Fig.~\ref{fig1}(b)). However, the associated noise from the thermal cloud interaction will also cause diffusive dynamics of the vortex (Fig.~\ref{fig1}(c)). Within the c-field formalism it is not correct to solely attribute the dissipative and noisy features of condensate evolution to the incoherent region, indeed the coherent region itself is a multi-mode system with its own intrinsic dissipative and fluctuation characteristics. Thus the relationship between the important processes and the various theories is not immediately clear.
The results of these distinct formalisms reveal that the SPGPE predicts the shortest vortex lifetime by a significant factor, suggesting that the other approaches neglect important physics and are not useful quantitative models.
 
Having established the importance of the method, we apply the full SPGPE to the study of vortex decay for an experimentally realistic system over the temperature range $0.78T_c\leq T\leq0.93T_c$, for traps with spherical and oblate geometries.  We present a technique for effectively estimating the SPGPE simulation parameters to obtain initial (equilibrium) states with a fixed total number as the temperature is varied.  We define a  vortex lifetime from the vortex decay trajectories, and characterize its variation with increasing temperature, finding a linear decline in the regime studied.   A comparison of vortex lifetimes for two geometries suggests that vortex bending modes associated with Kelvin waves provides an important mechanism of vortex decay that is controllable through the degree of trap oblateness. 

The angular momentum dynamics indicate regimes of both dissipative and diffusive decay of the vortex, where the latter regime is a consequence of thermal fluctuations from classical and quantum modes. The dissipative and diffusive vortex decay mechanisms are illustrated schematically in Fig~\ref{fig1}.

\section{Stochastic PGPE theory and models}
The derivation of the SPGPE for high temperature Bose gases has been given elsewhere~\cite{SGPEI,*SGPEII}. Here we briefly summarize our approach~\cite{Blakie08a,Weiler08a}.
A dilute Bose gas in the cold-collision regime is described by the second-quantized many body Hamiltonian
\EQ{\label{fullHamil}
H&=&\myint{\mbf{x}}\psi^\dag(\mbf{x})H_{\rm sp}\psi(\mbf{x})\nonumber\\
&&+\frac{u}{2}\myint{\mbf{x}}\psi^\dag(\mbf{x})\psi^\dag(\mbf{x})\psi(\mbf{x})\psi(\mbf{x}),
}
where the single-particle Hamiltonian and trap potential are
\EQ{
H_{\rm sp}&=&-\frac{\hbar^2\nabla^2}{2m}+V(\mbf{x}),\\
V(\mbf{x})&=&\frac{m}{2}\left[\omega_r^2(x^2+y^2)+\omega_z^2z^2\right]
}
and $u=4\pi\hbar^2a/m$, with $m$ the atomic mass and $a$ the s-wave scattering length. 

\begin{figure}[!t]
\begin{center}
\includegraphics[width=\columnwidth]{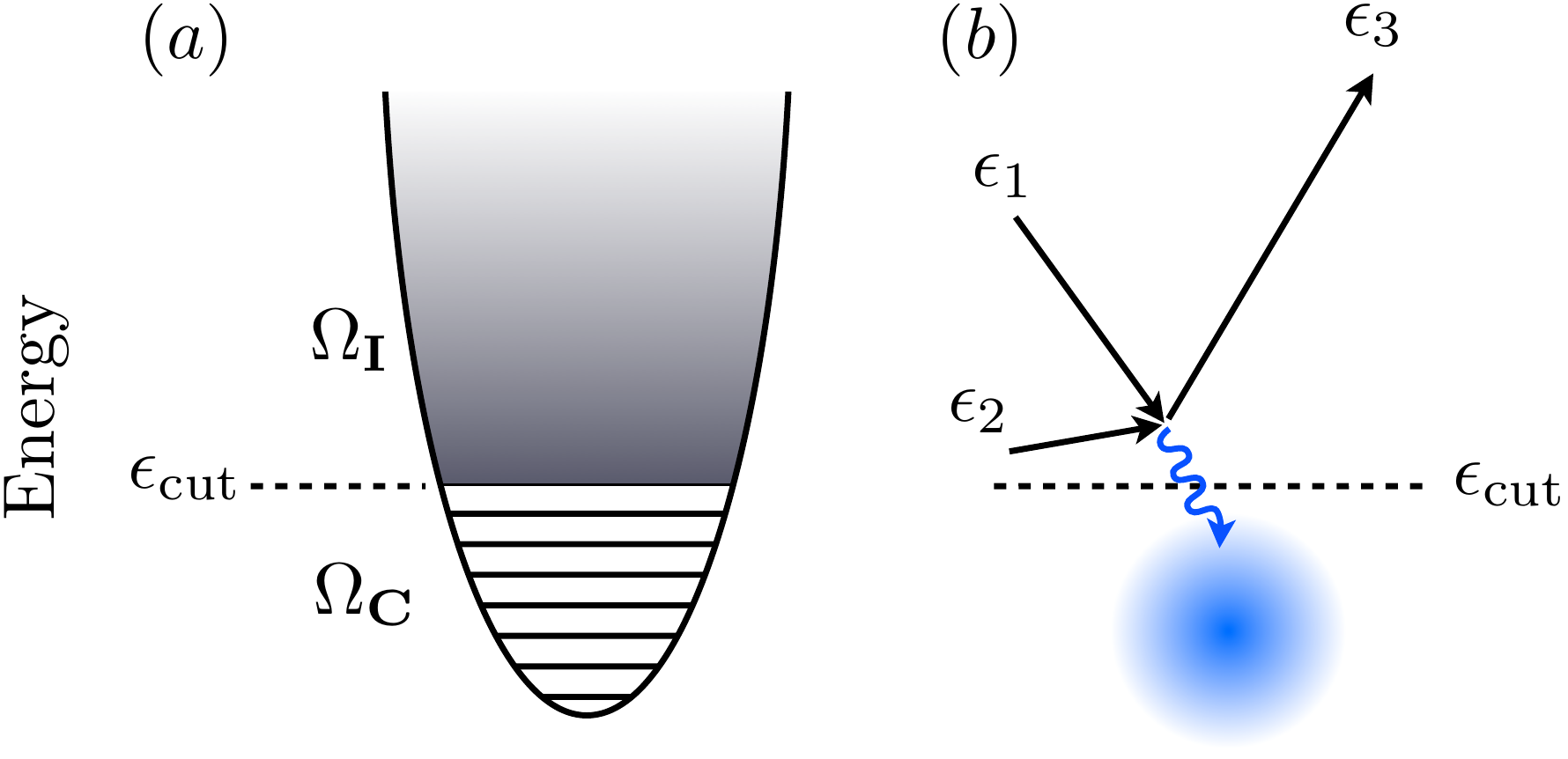}
\caption{(a) Separation of the high temperature Bose gas using the single particle energy cutoff $\ecut$, into coherent ($\rC$, with energies $\Omega_\rC\leq \ecut$) and incoherent ($\rI$, with energies $\Omega_\rI>\ecut$) regions. (b) The $\rI$ - $\rC$ collisions that lead to Bose-enhanced growth or loss: during a collision of two $\rI$ atoms, one atom absorbs most of the collision energy, allowing a stimulated transfer of the other into the coherent region, $\rC$. The time-reversed process also occurs.}
\label{fig2}
\end{center}
\end{figure}

As shown schematically in \fref{fig2}, the states of the trapped system are divided into what we refer to as the {\em coherent region} ($\rC$) of states with energy beneath the cut-off $\ecut$, and the remaining {\em incoherent region} ($\rI$) of high energy, quasi-equilibrium states.  The $\rC$ modes are assumed to be, at least moderately occupied and hence well-approximated by a classical field. This identification is key to the utility of the various c-field techniques as it allows the full coherent dynamics of the low energy portion of the system to be treated non-perturbatively.

The field operator is decomposed as
\EQ{\label{divide_field}
\psi(\mathbf{x})=\psi_\rC\of\x+\psi_\rI\of{\x}}
where $\psi_\rC\of\x=\PC\{\psi(\mathbf{x})\}$ and $\psi_\rI\of\x=\psi(\mathbf{x})-\psi_\rC\of\x$, are the field operators for the coherent and incoherent regions, respectively.
Our precise definition of the regions is made using a projection operator, $\PC$, defined by its action on a wave function  $f\of\x$ according to
\eqn{\label{PCdef2}
\PC\left\{f\of\x\right\}&=&\sum_{n\in \rC}\;\phi_n\of\x\myint{\xp}\phi_n^*\of\xp f(\xp),
}
where the $\phi_n\of\x$ are eigenstates of $H_{\rm sp}$, i.e.  $\epsilon_n\phi_n\of\x=H_{\rm sp}\phi_n\of\x$, and the restriction to the $\rC$ region in the summation is given by $\{n: \epsilon_n\le\ecut\}$. 

\subsection{Semi-classical treatment of the $\rI$ region}
In the SPGPE treatment the $\rI$ region modes act as a thermal reservoir for the low energy coherent region and are treated using master equation techniques.  
The situation we consider here, of a vortex damping from a partially condensed gas, is near equilibrium and it is adequate to assume that the $\rI$ region (which thermalizes rapidly compared to the $\rC$ region) is in equilibrium with a well-defined temperature $T$ and chemical potential $\mu$, and is well-described by a single particle Wigner function 
\begin{align}
F_\rI(\x,\mbf{k})&=\frac{1}{\exp{[(\hbar\omega(\mbf{x},\mbf{k})-\mu)/k_BT]}-1},\label{Fdef}
\end{align}
where  
\begin{equation}\label{Wdef}
\hbar\omega(\mbf{x},\mbf{k})=\frac{\hbar^2\mbf{k}^2}{2m}+V(\mbf{x}).
\end{equation}
By virtue of our cutoff this description is restricted to the $\rI$ region, and any quantities computed from (\ref{Fdef}) should only be integrated over the incoherent  region of phase space $\Lambda_\rI\equiv\{(\x,\mbf{k}):\hbar\omega(\mbf{x},\mbf{k})>\ecut\}$. For example, the number of \rI\ region atoms is given by 
\begin{equation}\label{eqn:nincomplete}
N_\rI=g_3\left(e^{\beta\mu},\beta\ecut\right)(\beta\hbar\bar{\omega})^{-3},
\end{equation}
where $\bar{\omega}\equiv(\omega_r^2\omega_z)^{1/3}$, and 
\begin{equation}
 g_\nu(z,y)\equiv\frac{1}{\Gamma(\nu)}\int_y^\infty dx\;x^{\nu-1}\sum_{l=1}^\infty(ze^{-x})^l,
 \end{equation}
is the incomplete Bose-Einstein function \cite{Bradley08a}. 

\subsection{SPGPE theory of the $\rC$ region}
By formally tracing over the $\rI$ region, we arrive at a  description for the dynamics of the $\rC$ region field in which the influence of the $\rI$ region appears as damping ($\gamma$) and complex noise  ($dW_\gamma$) terms. The evolution equation has the form
\begin{equation}
\begin{split}
d\psi_\rC(\mbf{x},t)=&\PC\Big\{ -\frac{i}{\hbar}L_\rC\psi_\rC(\mbf{x},t)dt\\
&+\frac{\gamma}{k_BT}(\mu-L_\rC)\psi_\rC(\mbf{x},t)dt+dW_\gamma(\mbf{x},t)\Big\},
\label{SGPEsimp}
\end{split}
\end{equation}
which we refer to as the simple growth SPGPE.
The operator $L_\rC$ is the Hamiltonian evolution operator for the $\rC$ region 
\EQ{\label{LCdef}
L_\rC\psi_\rC(\x)\equiv \left(H_{\rm sp}+u|\psi_\rC(\x,t)|^2\right)\psi_\rC(\x),
}
and the complex noises are given by \begin{eqnarray}
\label{GamCor}
\langle dW_\gamma^*(\mbf{x},t)dW_\gamma(\mbf{x}^\prime,t)\rangle&=&2\gamma\delta_C(\mbf{x},\mbf{x}^\prime)dt,\\
\langle dW_\gamma(\mbf{x},t)dW_\gamma(\mbf{x}^\prime,t)\rangle&=&\langle dW_\gamma^*(\mbf{x},t)dW_\gamma^*(\mbf{x}^\prime,t)\rangle=0,\;\;\;\;
\end{eqnarray} 
where $\delta_\rC(\x,\xp)=\sum_{n\in \rC}\phi_n\of\x \phi_n^*\of\xp$ is a delta-function in $\rC$. This result is systematically derived in Ref.~\cite{Bradley08a}) using the truncated Wigner approximation and master equation techniques, with the notable feature that the damping and noise parameters are calculable from our description of the $\rI$ region for a system close to thermal equilibrium, giving the expression
\begin{eqnarray}\label{gamdef} \gamma&=&\gamma_0\Bigg\{\left[\ln{\left(1-e^{\beta(\mu-\ecut)}\right)}\right]^2 +e^{2\beta(\mu-\ecut)} \times \nonumber \\
&&\sum_{r=1}^\infty\;e^{r\beta(\mu-2\ecut)}\Phi[e^{\beta(\mu-\ecut)},1,r+1]^2\Bigg\},\;\;\;\;\;\;\;\;\;
\end{eqnarray}
where $\gamma_0= 4m(ak_BT)^2/\pi\hbar^3$, and $\Phi[x,y,z]$ is the Lerch transcendent.  

The second line of the SPGPE \eref{SGPEsimp} is directly responsible for condensate growth from scattering between two \rI\ region atoms, as shown schematically in \fref{fig2}(b). $\mu$ and $T$ are respectively the chemical potential and temperature of the thermal reservoir of particles in the \rI\ region, defined in Eq.~(\ref{Fdef}). By using the spatially invariant form (\ref{gamdef}) for the damping, we have neglected a weak spatial variation arising at the edge of the $\rC$ region~\cite{Bradley08a}.

Additional terms from the full SPGPE theory that represent number conserving scattering processes between atoms in \rC\ and \rI\ \cite{SGPEII} have also been omitted. These processes do not lead to condensate growth, and are weak in near equilibrium situations~\cite{Blakie08a}. The numerical implementation of Eq.~(\ref{SGPEsimp}) is relatively straightforward, with individual trajectories being only slightly more computationally intensive than integration of the PGPE. Our method of numerical stochastic integration has been given in a number of places; we refer the reader to Refs.~\cite{SGPEI,*SGPEII,Bradley08a,Blakie08a} for further details.

We note the related stochastic GPE approach of Stoof and coworkers \cite{Stoof1999,*Stoof2001,*Duine2001,*Proukakis06a,*Cockburn09a,Duine2004} (also see \cite{Damski09a}), which differs from our approach most significantly through the absence of an explicit projector. A comparison of these approaches can be found in Ref.~\cite{Proukakis08a}. It is our opinion that the projector is a crucial element of the reservoir theory that cannot be simply chosen arbitrarily, as is extensively discussed in Ref.~\cite{Blakie08a}.

The SPGPE description provides a grand canonical description of the $\rC$ region so that, irrespective of the value of $\gamma$, in equilibrium it samples microstates with probability $P(\psi_\rC)\propto\exp\left(-K_\rC/k_BT\right)$, where $K_\rC\equiv H_\rC-\mu N_\rC$ is the grand-canonical Hamiltonian  and
\begin{align}
H_{\rC}&=\int d^3\x\,\psi_\rC^*(\mbf{x})\left(H_{\rm{sp}}+\frac{u}{2}|\psi_\rC(\mbf{x})|^2\right)\psi_\rC(\mbf{x}),\label{Hc}\\
N_\rC&=\int d^3\x\,|\psi_\rC(\mbf{x})|^2\label{Nc},
\end{align}
are the expressions for the  c-field energy and number respectively. We now consider two simplified theories that can be obtained from the SPGPE by neglecting various effects.

\subsection{Damped PGPE  treatment}
A damped PGPE (DPGPE) is obtained by neglecting the noise term in the SPGPE (\ref{SGPEsimp}), and is of the form
\begin{equation}
\begin{split}
i\hbar \frac{\partial \psi_\rC(\mbf{x},t)}{\partial t}=&\PC\left\{L_\rC\psi_\rC(\mbf{x},t)-i\frac{\hbar\gamma}{k_BT}(L_\rC-\mu)\psi_\rC(\mbf{x},t)\right\}.
\label{PGPEdamp}
\end{split}
\end{equation}
While the SPGPE serves to minimize the grand free energy of the $\rC$ region, evolution according to \eeref{PGPEdamp} instead minimises $K_\rC$:
\EQ{\label{Kc}
\frac{dK_\rC}{dt}=-\frac{2\gamma}{k_BT}\intV{\x}|\PC\left\{(\mu-L_\rC)\psi_\rC(\x)\right\}|^2,
}
i.e. thermal fluctuations are damped out, and the equilibrium solution is the zero temperature ground state, satisfying $\PC L_\rC\psi_\rC=\mu\psi_\rC$. An initial state containing a vortex will decay to a vortex-free state (provided the  $\rI$-region is rotating sufficiently slowly), on a timescale set by $\gamma$.

To date several numerical studies of vortex nucleation and dynamics have been based upon a phenomenological damped Gross-Pitaevskii equation~\cite{Penckwitt2002,Tsubota2002,Kasamatsu2003}. The description was first introduced in \cite{Tsubota2002} on the grounds that it gives the correct equilibrium solution, and that it follows naturally from the damped GPE approach of~\cite{Choi1998} for a cloud in equilibrium at rotation frequency $\Omega$. This approach received further justification via a stochastic treatment of the dissipative interaction of a BEC with a rotating thermal cloud~\cite{SGPEI,Bradley08a}.
As shown by Kasamatsu \etal~\cite{Kasamatsu2003}, the equation of motion may also be obtained from the generalized finite temperature GPE formalism of Zaremba, Nikuni, and Griffin (ZNG)~\cite{Zaremba1999}.

\subsection{PGPE treatment}
By setting $\gamma\to0$ (in addition to $dW_\gamma\equiv0$) we arrive at an isolated PGPE description of the $\rC$ region 
\begin{equation}
\begin{split}
i\hbar \frac{\partial \psi_\rC(\mbf{x},t)}{\partial t}=&\PC\left\{L_\rC\psi_\rC(\mbf{x},t)\right\}.
\label{PGPE}
\end{split}
\end{equation}
The PGPE is also formally obtained as Hamilton's equation of motion for the c-field Hamiltonian \eref{Hc}:
\EQ{
i\hbar\frac{\partial  \psi_\rC(\mbf{x})}{\partial t}=\frac{\bar{\delta}H_\rC}{\bar{\delta}\psi_\rC^*(\mbf{x})},
}
where the restriction to the $\rC$ region arises naturally in the projected functional differential formalism of Ref.~\cite{SGPEII}.
An interesting feature of the PGPE theory is the emergence of ergodicity from Hamiltonian evolution~\cite{Blakie08a} (particularly for the finite temperature regime we consider, where the $\rC$ region contains many appreciably occupied modes). While the PGPE approach may appear to be a simplification of the DPGPE, it provides quite a distinct physical description. It is a microcanonical model of the $\rC$ region in which both $H_\rC$ and $N_\rC$ are conserved and, by virtue of $\gamma$ being zero, the full complement of coherent region excitations (and their intrinsic noisy effect on the $\rC$ region dynamics) are retained.

Two recent applications of a rotating frame PGPE~\cite{Bradley08a} have studied vortex lattice formation~\cite{Wright08a}, and symmetry breaking associated with the single vortex in equilibrium~\cite{Wright09a}. Other treatments of the isolated coherent region have been performed, typically without a projector and have adopted the \textit{classical field} moniker. Important earlier works are those of Schmidt \etal~\cite{Schmidt2003}, who considered the decay of a vortex in a purely Hamiltonian GPE evolution of a \emph{thermal} Bose field containing a vortex, and of Parker and Adams~\cite{Parker05a,Parker06a} who studied the formation of a vortex lattice in a stirred, zero temperature BEC. 

\section{State preparation and vortex detection}\label{sec:sgpeeq}
In our simulations the same initial conditions are used for the SPGPE, DPGPE and PGPE calculations, and are all prepared using the SPGPE to obtain an finite temperature equilibrium state. A central vortex is then generated in the state, and its position is detected as a function of time. In this section we discuss the procedure used to generate these states and distinguish the vortex from surface fluctuations.

\subsection{Efficient scheme for preparing initial states at fixed $N_{\rm T}$}
 In many applications it is desirable to perform simulations in which the temperature is changed but the total atom number in the system, $N_{\rm T}$ (and  hence critical temperature $T_{\rm{c}}$), remains constant. Equilibrium states in the SPGPE description are generated by evolving Eq.~(\ref{SGPEsimp}) with choices for $T$ and $\mu$, but as the relationship between $\mu$ and $N_{\rm T}$ is nonlinear, obtaining the desired $N_{\rm T}$ (which can only be found a posteriori) can require running a large number of simulations to determine the appropriate $\mu$ value by trial and error.
Another parameter that must be determined is the value of $\ecut$ used to define the $\rC$ region. The validity of the SPGPE approach requires that the highest mode in the $\rC$ region be appreciably occupied (usually with a occupation in the range $\ncut=1-5$) and to ensure this value of $\ecut$ will also change with $T$ where necessary (e.g.~ see \cite{Blakie2007a}).  Because the projection is made using the single particle basis, we also require a sufficiently large cutoff to describe the interacting modes of the system. In practice this requirement is  satisfied if $\ecut\gtrsim2\mu$.

Here we develop a simple mean field method that provides useful estimates for the choices of the parameters $\mu$ and $\ecut$ for a system of $N$ atoms at temperature $T$.  

In our approach the condensate is treated within in the Thomas Fermi approximation, with the number of condensate atoms ($N_0$) determining the system chemical potential. The thermal atoms are then treated using  the semi-classical Hartree-Fock approximation, i.e.~with energies 
\eqn{\label{HFH}
E_{\rm HF}(\x,\p)\equiv \frac{\p^2}{2m} + V(\mbf{x} ) + 2u n_0(\mbf{x}) 
}
where $n_0(\mbf{x})$ is the condensate (Thomas-Fermi) density.  The Hartree-Fock density of states can not in general be obtained analytically, but it can easily be handled for the case of a harmonically trapped condensate, as shown in Ref.~\cite{Bijlsma00a}. 
Following this approach we have obtained an efficient numerical algorithm for estimating $\mu$ and $\ecut$ at a given temperature and total atom number. The details of this procedure are given in Appendix \ref{DOSdetails}.
\begin{figure}[t!]
\begin{center}
\includegraphics[width=\columnwidth]{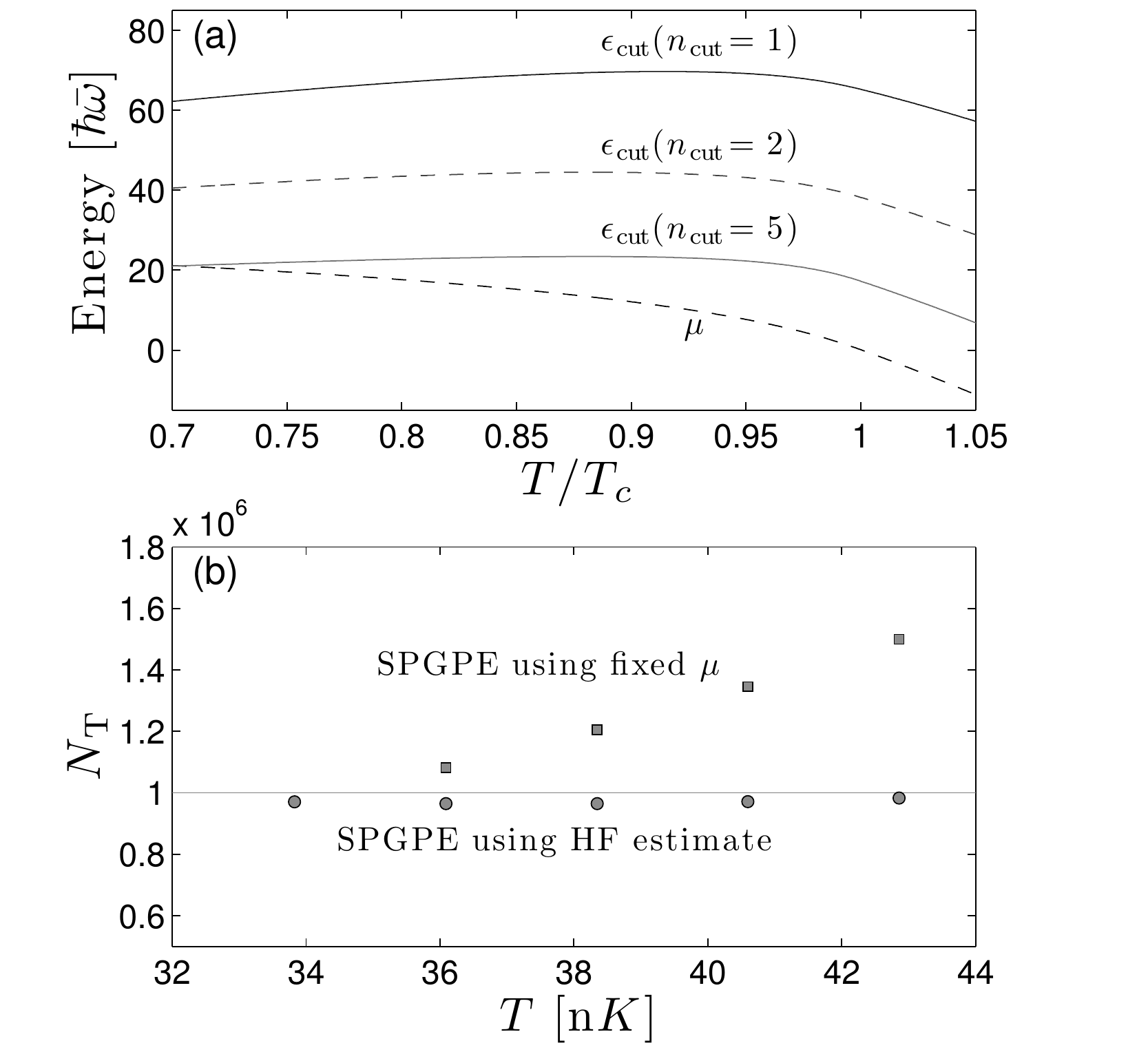}
\caption{Estimated and SPGPE parameters as a function of input temperature using an input value of  $N_{\rm T}=1 \times 10^6$ for a spherical system with trap frequency $\omega = 2\pi \times 10 \ s^{-1}$. (a) Hartree-Fock estimates of the cutoff energy $\ecut$ for a range of cutoff mode occupations $\ncut$. The dashed lines give the curves for $\ecut$ and $\mu$ used in (b) Total atom number from SPGPE simulations evolved to equilibrium using HF estimated parameters (circles), and fixed $\mu$ (squares). $\mu$ is held at the value giving $N_{\rm T}$ at $T=34$ nK. The solid line is the target value of $N_{\rm T}$.}
\label{fig3}
\end{center}
\end{figure}

Results from the Hatree-Fock calculation for $\mu$ and $\ecut$ for a system of $1\times 10^6$ $^{87}$Rb atoms are shown in 
Fig.~\ref{fig3} (a), revealing the typical way in which these parameters need to change with temperature to ensure fixed total number.  We see that the value of $\ecut$ for  $\ncut=2$ is sufficiently large that the requirement $\ecut\gtrsim2\mu$ is well-satisfied in the  temperature regime $0.7 T_{\rm{c}} < T  \sim T_{\rm{c}}$. It should be noted that the Hartree-Fock predictions only depend  on the trapping frequencies through $\bar{\omega}$, so that these parameters can be applied to any system of equal geometric mean, provided a three dimensional description is appropriate.

Because of the simplicity of the Hartree-Fock theory it is necessary to test that its predictions are accurate.
In what follows we present SPGPE results using the Hartree-Fock estimates for   $\mu$ and $\ecut$ to verify the accuracy at which the desired total number is obtained for various temperatures. 
To do this for each set of parameters we determine the incoherent region population, $N_\rI$, according to  Eq.~(\ref{eqn:nincomplete}), and from the equilibrium state obtained from the SPGPE simulation using the value of $\mu$ and $\ecut$ predicted by Hartree-Fock theory, we determine $N_\rC$, and thus the total number $N_{\rm T}=N_\rC+N_\rI$.
\begin{figure*}[tpb!]
\begin{center}
\includegraphics[width=\textwidth]{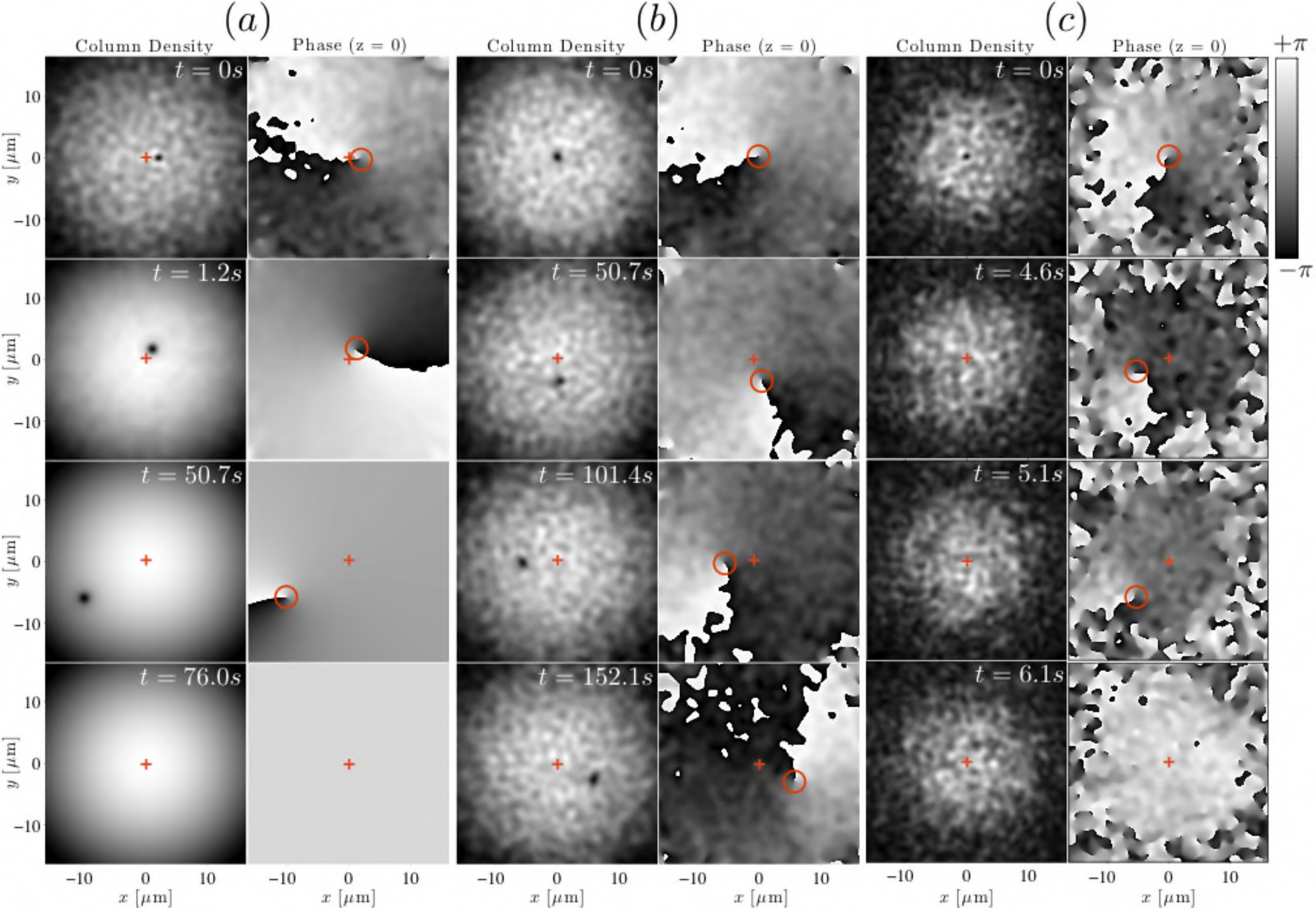}
\caption{Vortex dynamics using the DPGPE and PGPE in a system with a 4:1 geometry.  The vortex is most easily observed from the phase singularity (circle), and the center of the condensate is marked by a cross (+).  (a): DPGPE vortex evolution for $T/T_{\rm{c}} = 0.78$, where the vortex has an initial offset from the origin of the condensate. The initial noise is damped out  within $\sim1.2$s, after which the vortex decay is governed by $\gamma$.  (b): PGPE vortex evolution for $T/T_{\rm{c}} = 0.78$, with an initially centered vortex. Despite initial thermal fluctuations kicking the vortex off the center, the size of the thermal component is too small at this temperature to see a finite lifetime.  (c): PGPE vortex evolution for a system at $T/T_{\rm{c}} = 0.93$ where the initially centered vortex has a finite lifetime due to increased classical fluctuations at this temperature.}
\label{fig4}
\end{center}
\end{figure*}
To obtain equilibrium states, the SPGPE [Eq.~(\ref{SGPEsimp})] is evolved with the parameters determined using the Hartree-Fock scheme. Note that the grand canonical equilibrium solution is independent of the size of $\gamma$, allowing us to find equilibria more rapidly using a dimensionless damping $\hbar\gamma/K_BT\sim 0.3$ that is much larger than typical dynamical values ($\hbar\gamma/K_BT\sim 10^{-3}$). We then find that the c-field reaches thermal equilibrium after propagation for much less than a single radial trap period, when initialized from a Thomas-Fermi initial state. 

Our results are calculated for a Bose gas in a spherical trap with frequency $\omega = 2 \pi \times 10 \ $s$^{-1}$ and  $N_{\rm T} = 1 \times 10^6$ atoms.  Figure \ref{fig3} (b) shows the SPGPE results for $N_{\rm T}$ over a range of temperatures and reveals that the HF estimated parameters generate equilibrium SPGPE solutions with total atom number in close agreement for the desired number.
  For comparison we also show the effect of holding $\mu$ constant and increasing temperature in Fig.~\ref{fig3} (b). In this case the total number of atoms increases quite rapidly, and hence the critical temperature also increases ($k_{\rm{B}} T_c = 0.94 \hbar \bar{\omega} N^{1/3}).$
\begin{table}[t]
\begin{center}
\begin{tabular}{c  c  c }
\hline\hline$T / T_{\rm{c}}$ (input) \hspace{.2cm} &\hspace{.2cm} constant $\mu$\hspace{.2cm} &\hspace{.2cm} HF-estimated $\mu$\\
\hline
\rule{0pt}{1ex} 0.75 & 0.76 & 0.76  \\
 
\rule{0pt}{1ex} 0.80 & 0.78 & 0.81  \\

\rule{0pt}{1ex} 0.85 & 0.80 & 0.86  \\

\rule{0pt}{1ex} 0.90 & 0.82 & 0.91  \\

\rule{0pt}{1ex} 0.95 & 0.83 & 0.96  \\
\hline\hline
\end{tabular}
\caption{Comparison of input and resulting SPGPE reduced temperatures for a spherical  system with $\omega_r = 2\pi \times 10  {\rm s}^{-1}$ and input atom number $N_{\rm T} = 1 \times 10^6$.}
\label{table temp}
\end{center}
\end{table}
Table \ref{table temp} compares the two methods, for the results of Fig.~\ref{fig4}. The SPGPE results using the HF parameter estimates effectively and accurately span the temperature range where the SPGPE is valid. In comparison, the fixed $\mu$ case approaches the transition slowly since $N_{\rm T}$ increases. We have also verified that the method works over a range of geometries. 
The c-field occupation at the cutoff is typically slightly smaller than the input value. For example, we find that using an input value of $\ncut=2$ to the HF-estimate of $\ecut$ results in SPGPE solutions with $\ncut\sim 1.5$. 

\subsection{Preparing a vortex state}\label{makeV}
Finite temperature states containing a vortex line that lies on the cylindrical symmetry axis of the trap ($z$), are induced by imposing a vortex phase pattern on samples of the SPGPE equilibrium state, $\psi_\rC(\x)\to \psi_\rC(\x)e^{i\phi(\x)}$, where
\eqn{
\phi(\x)=\arctan{\left(\frac{y}{x}\right)}.
}
Imposing a phase consistent with a single quanta of circulation has the effect of only coupling to the coherent region, so that the thermal fraction above $\ecut$ and the high energy thermally occupied modes of the c-field region are not changed. Thus the system is initialized as a finite temperature equilibrium state with non-rotating thermal component, containing a central, axially aligned, singly charged vortex.
\begin{figure*}[tpb]
\begin{center}
\includegraphics[width=.8\textwidth]{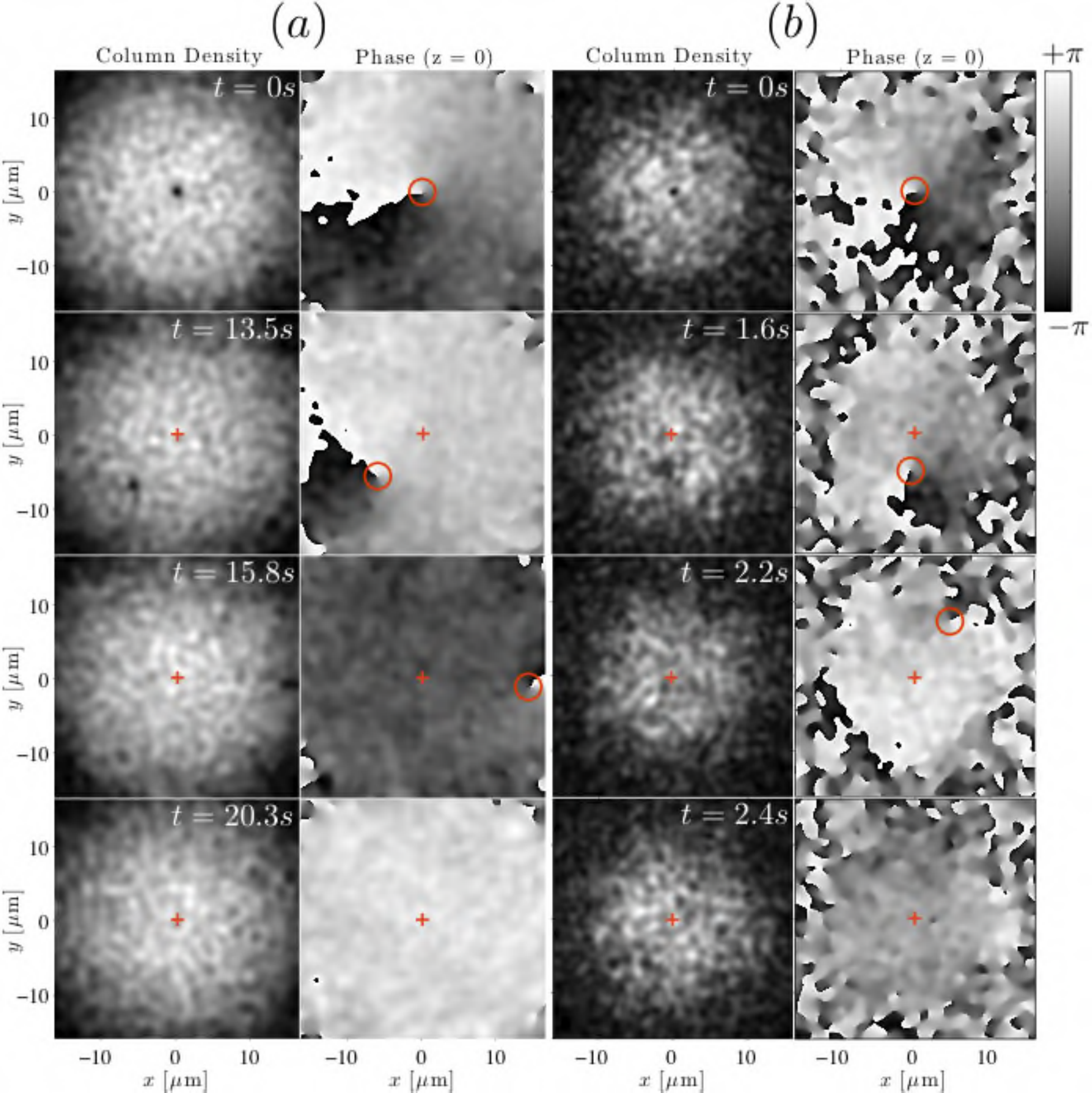}
\caption{Column densities and phase slices showing SPGPE vortex evolution in a system with aspect ratio of 4:1 at temperatures (a) $T/T_{\rm{c}} = 0.78$, and (b) $T/T_{\rm{c}} = 0.93$. The vortex is most easily identified from the phase singularity (circle), and the center of the condensate is marked by a cross (+). Other parameters are for a gas of $5\times 10^5$ $^{87}$Rb atoms in a harmonic trap with $(\omega_r,\omega_z)=2\pi(12.4,49.7){\rm Hz}$.}
\label{fig5}
\end{center}
\end{figure*}
\subsection{Vortex detection}
To unambiguously detect the a vortex as it migrates to the condensate edge, we use an adaptive method that compares the vortex signal with surface fluctuations. In detail, the curl of the velocity field $\mathbf{\omega}(\x)=\nabla\times \mathbf{v}(\x)$, is used as our basic signal. Taking a slice through the c-field at $z=0$, a central vortex generates a significant signal localized at the vortex core, which is unambiguously resolvable; we denote the vortex displacement from the $z$-axis by $r_v$. However, near the condensate boundary, the vortex is more difficult to distinguish from phase singularities in the evanescent matter field caused by the stochastic noise term in the SPGPE. In experiments this would correspond to a loss of core visibility in absorption images due to thermal fluctuations. We thus introduce a radial mask to remove surface fluctuations from the detection process by determining the maximum penetration depth of surface fluctuation phase singularities for a given temperature; this is accomplished by an adaptive routine that is applied to the c-field data for each trajectory. 

At short times, the surface excitations are clearly distinguishable from the central vortex. Our numerical algorithm for each trajectory is as follows: (a) at each time step we detect all vortices with radius $r_v<R_v$. (b) we discard any that are further than a small numerical tolerance from the location of the decaying vortex at the previous time step. $R_v$ is then adjusted inwards to the smallest radius of any spurious vortex signal. The mask radius moves rapidly inward until it eliminates all surface fluctuations from the vortex detection region. We take this to be the largest vortex core displacement radius at which the vortex is still resolvable, $R_v$, and use it to define the vortex lifetime in a single SPGPE trajectory $t_v\equiv {\rm max}(t : r_v<R_v)$.The specific value of $R_v$ is temperature, and, less directly, number dependent. At zero temperature $R_v\simeq R_{\rm TF}$; for $T\to T_c$, $R_v\to 0$, demonstrating the shrinking phase coherent region inside the BEC. As the SPGPE is a stochastic theory, in what follows we compute the \textit{mean vortex lifetime}, $\bar{t}\equiv\langle t_v\rangle$, over a set of trajectories for each temperature considered. For simplicity, we hereafter refer to the trajectory-averaged $\bar{t}$ as the vortex lifetime. We find that this procedure gives a time for vortex death that corresponds to a specific value of angular momentum per particle ($\sim 0.3\hbar$) that is \emph{independent} of the system temperature. 
\section{Results}

To understand the different decay mechanisms we first compare the dynamics of a centrally located vortex by studying single trajectories within PGPE, DPGPE, and SPGPE theories. 
\subsection{Comparison of models of vortex decay}
Using the initial finite $T$ central vortex states (see \sref{makeV}), we compare subsequent evolution with respect to the three approaches. For the PGPE case, we neglect all interactions with the above cutoff region in the SPGPE. For the DPGPE case we neglect only the noise, and we also introduce a small initial offset of the vortex. The offset is necessary because a central vortex is metastable, and hence evolution according to the purely damped PGPE will not exhibit vortex decay unless there is sufficient symmetry breaking. Initial fluctuations and noise in the PGPE and SPGPE approaches provide sufficient symmetry breaking to cause decay, but the initial fluctuations in the DPGPE treatment are insufficient. For the DPGPE and SPGPE, cases the value of $\gamma$ is determined by Eq.~(\ref{gamdef}). 

A comparison of individual trajectories for the DPGPE and PGPE approaches is presented in Figure~\ref{fig4} for a system in a 4:1 trapping geometry with $(\omega_r,\omega_z)=2\pi\times(12.4,49.7){\rm Hz}$. In Fig.~\ref{fig4} (a) and (b) the initial temperature is $T=34.0\;{\rm nK}$. In the DPGPE approach shown in Fig.~\ref{fig4}(a), initial thermal fluctuations are quickly damped out, the field density becomes smooth, and the vortex then decays according to familiar damped GP evolution. The small value of $\hbar\gamma/k_BT=7.1\times 10^{-4}$ leads to a long lifetime of $\bar{t}= 69.4$s. Figure~\ref{fig4}(b) shows PGPE evolution for the same system. The vortex evolves into a quasi-equilibrium state of precession about the $z$-axis at an approximately fixed radius. At this value of temperature, the angular momentum in the initial vortex configuration is too large to be completely transferred to the non-condensate fraction of the classical field, inhibiting vortex decay. 

In Fig.~\ref{fig4}(c) the initial state has temperature $T=64.9$~nK and thus contains a significantly larger non-condensate fraction. After a short time (3.5s), the vortex irreversibly leaves the condensate through exchange of angular momentum with non-condsate degrees of freedom.
\begin{table}[!t]
\begin{center}
\begin{tabular}{c|ccc }
\hline\hline \hspace{.2cm}$T/T_c$\;\;& \hspace{.2cm}Theory\hspace{.2cm} & \hspace{.2cm} $\bar{t}$ (sec) \textbf{1:1}\hspace{.2cm}  &\hspace{.2cm} $\bar{t}$ (sec) \textbf{4:1} \\
\hline
\multirow{3}{*}{0.78}&PGPE & $\infty$ & $\infty$\\
&DPGPE & 33.7 & 69.4\\
&SPGPE & 10.7 & 20.4\\ \hline
\multirow{3}{*}{0.93}&PGPE & 2.6 & 3.5\\
&DPGPE & 21.0 & 53.9\\
&SPGPE & 1.5 & 2.0\\ \hline\hline
\end{tabular}
\caption{Comparison of vortex lifetime in different theories for a vortex in a Bose gas with spherical (1:1) and oblate (4:1) geometries.}
\label{tableTh}
\end{center}
\end{table}
Figure~\ref{fig5}(a) shows SPGPE trajectory results for the same parameters as Fig.~\ref{fig4}(a), (b). While the vortex did not decay in PGPE and decayed slowly in DPGPE, the combined influence of classical fluctuations in the c-field and reservoir interactions leads to a much shorter lifetime of $\sim 20$s in the 4:1 geometry. Figure~\ref{fig5}(b) shows an SPGPE trajectory for the parameters of Fig.~\ref{fig4}(c). The SPGPE lifetime of 2s is significantly shorter than the 3.5s lifetime of the PGPE, which is shorter than the DPGPE lifetime of 54s. The results for lifetimes are summarized in Table~\ref{tableTh}, where we also show lifetimes for a spherical trap. We note that the temperature dependence of the lifetime of a vortex evolved under the DPGPE is much weaker than under SPGPE evolution; the SPGPE lifetime changes by an order of magnitude between the two temperatures considered.

While we have not yet performed a systematic study of the influence of trap geometry, we note that the main effect of reducing the oblateness of the trap (note that $\bar{\omega}$ is held constant) appears to be to shorten the lifetime further. Presumably this is associated with the accessibility of vortex bending modes or Kelvin waves~\cite{Fetter04a,*Simula08a} which provide a further dissipative mechanism that is not available in highly oblate configurations.
\begin{figure}[!t]
\begin{center}
\includegraphics[width=\columnwidth]{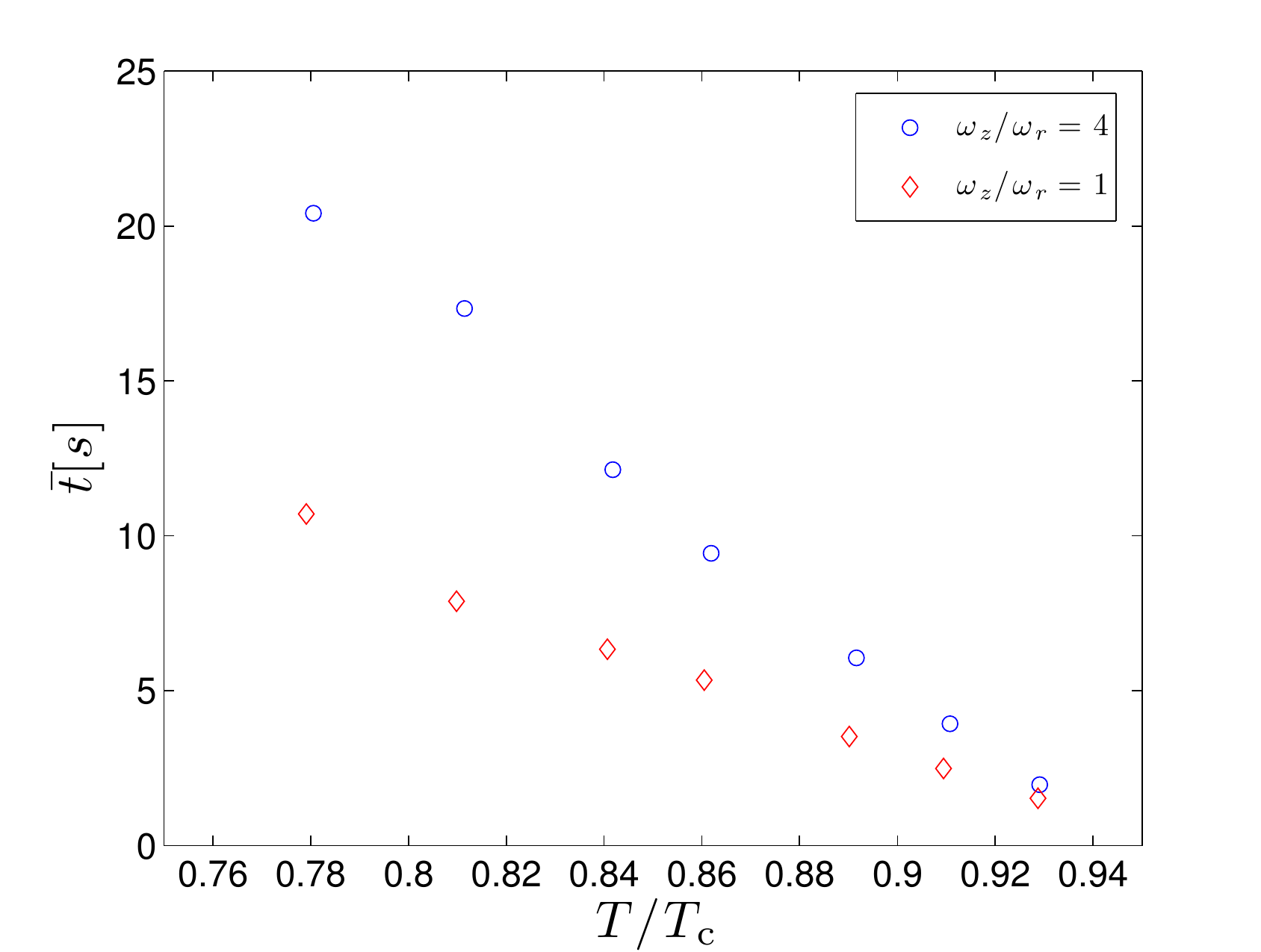}
\caption{Lifetime of a single vortex, $\bar{t}$, as a function of temperature for a system containing $5\times 10^5$ $^{87}$Rb atoms. The trap has fixed $\bar{\omega}=2\pi\times 19.7\;{\rm s^{-1}}$, so that $T_c=69.7$~nK is the same for both geometries.}
\label{fig6}
\end{center}
\end{figure}
\subsection{SPGPE simulations of vortex decay}
We perform simulations of the SPGPE for a range of temperatures and calculate the vortex lifetime $\bar{t}$. All quantities of interest are averaged over an ensemble of trajectories which is chosen sufficiently large to give an acceptable level of stochastic convergence. For each temperature we choose parameters in order to keep $N_{\rm T}=5\times 10^5$ constant, using the estimation procedure of Sec.~\ref{sec:sgpeeq}. The resulting $\mu$ is used to evaluate Eq.~(\ref{gamdef}) to determine the strength of dissipation and noise in the SPGPE.

The mean vortex lifetime for 4:1 and 1:1 geometries is shown in Fig.~\ref{fig6}. For both geometries the geometric mean  and temperature are $\bar{\omega}=2\pi\times 19.7\;{\rm s}^{-1}$, and $T_c=69.7\;{\rm nK}$ respectively. Rather surprisingly, we observe that the vortex lifetime is linear in $T/T_c$ at fixed $N_{\rm T}$. This result may not be totally unexpected since the leading dependence of the damping on temperature is given by the prefactor of the dimensionless damping rate $\hbar\gamma/k_BT\sim 4ma^2k_BT/\pi\hbar^2$. However, for constant $N_{\rm T}$ the temperature variation leads to large changes in chemical potential, and thus nonlinear changes in $\gamma$~\cite{Blakie08a}. The characteristic size of the condensate also shrinks with increasing $T$ [decreasing $\mu$, see Fig.~\ref{fig3} (a)], so the vortex is not only more mobile, it has a shorter distance to traverse (e.g. the condensate radius in Fig.~\ref{fig5}(a) is $\sim 15\mu{\rm m}$, while in Fig.~\ref{fig5} (b) it has reduced to $\sim10\mu{\rm m}$). We note that the slope of $\bar{t}$ increases with $\omega_z/\omega_r$. We again associate this increased dissipation with the accessibility of vortex bending modes for the more spherical system. This connection can be made because we have held all parameters of the system constant, apart from temperature and the ratio $\omega_z/\omega_r$. Spurious factors  that would change the vortex decay in an uncontrolled way, such as changing $N_{\rm T}$ or $\bar{\omega}$, have been eliminated.

\begin{figure}[t]
\begin{center}
\includegraphics[width=\columnwidth]{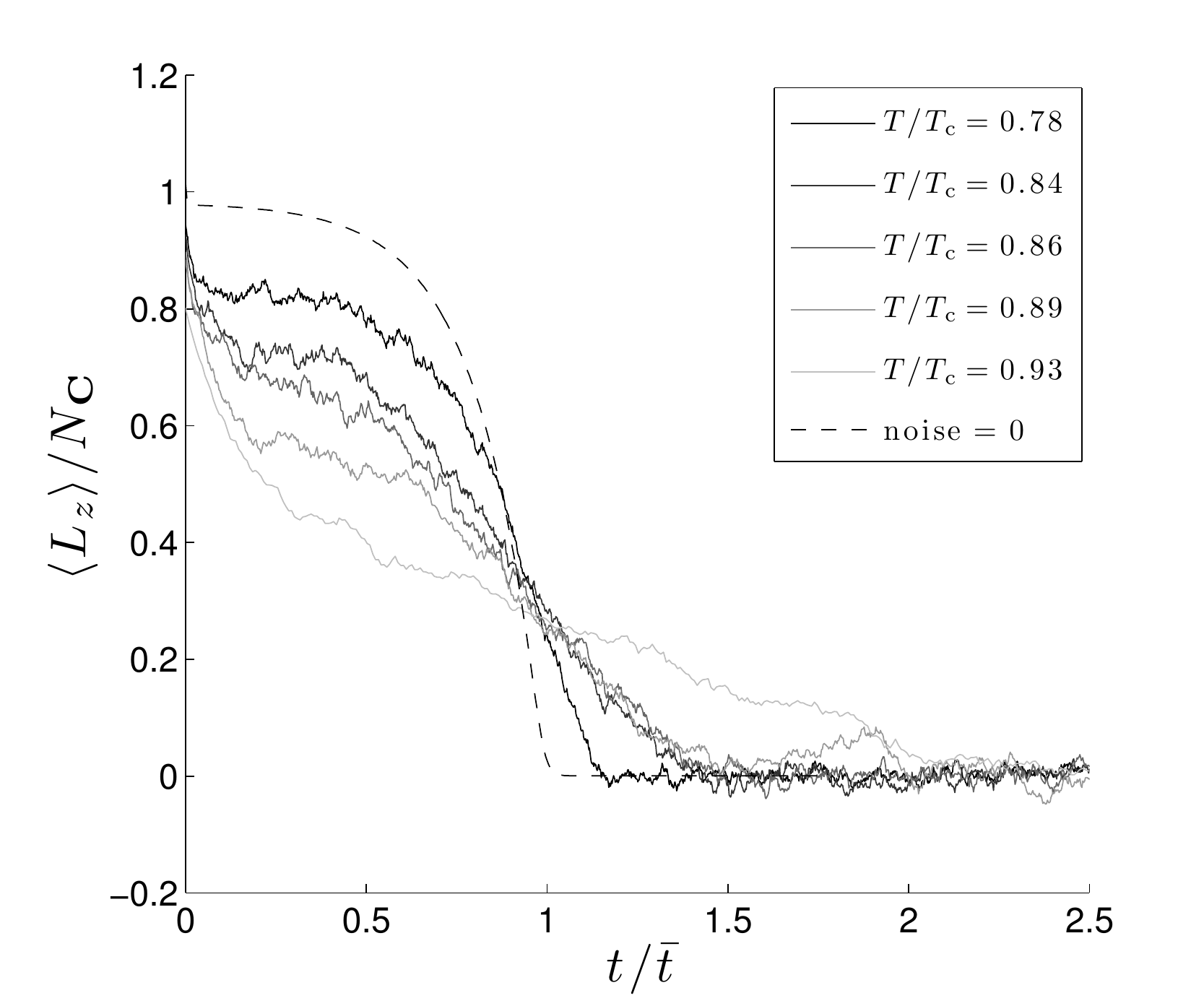}
\caption{The $z$ component of the averaged angular momentum per particle as a function of $t/\bar{t}$, where $\bar{t}$ is the vortex lifetime, for a system with an aspect ratio of 4:1 as found from an ensemble average of SPGPE trajectories.  The case of purely dissipative decay is given by the dotted line. Note the wide contrast in the form of the vortex decay where for $T/T_{\rm{c}} = 0.78$ we see dissipation dominating the decay, while for higher temperatures diffusive decay becomes more prevalent. }
\label{fig7}
\end{center}
\end{figure}

Figure~\ref{fig7} shows the angular momentum per particle of the $\rC$ region during vortex decay. To identify regimes of different qualitative behavior we graph the data in units of the vortex lifetime $\bar{t}$ calculated for each temperature. The point $t=\bar{t}$ nearly coincides with the intersection point of all curves, namely where $\langle L_z\rangle/N_\rC\approx 0.3$. This corresponds to the angular momentum remaining when the vortex is indistinguishable from thermal fluctuations at the condensate boundary. We also plot $\langle L_z\rangle/N_\rC$ for the overdamped limit of the DPGPE, which drops to zero at $t=\bar{t}$; this corresponds to the time when the vortex reaches the Thomas-Fermi radius of the BEC. Comparing with the overdamped limit, we see that for low temperatures, the form of the decay in the SPGPE approaches that of purely dissipative decay governed by the DPGPE. The main effect of the additional thermal fluctuations in the SPGPE is to cause rapid loss of angular momentum at short times, corresponding to destabilizing the vortex from its initial position at the origin. Thus, thermal fluctuations render the vortex metastable. 

As temperature increases, the angular momentum becomes increasingly persistent for $\bar{t}<t$. The angular momentum develops a broad tail at long times associated with a broader distribution of vortex radii. We refer to this high temperature behavior as the regime of \emph{diffusive decay}, where the physics of vortex decay is dominated by the diffusive influence of thermal fluctuations. 
\section{Conclusions}
We have presented a scheme to conveniently estimate consistent parameters in the regime of validity of the SPGPE for systems near thermal equilibrium. The chemical potential is calculated using a Hartree-Fock calculation of the $\rC$-region occupation. The cutoff energy $\ecut$ is an essential feature of the theory used to consistently separate the coherent and incoherent regions of the system. It is determined by finding the single particle cutoff that bounds the same number of quantum states as contained in the Hartree-Fock treatment of $\rC$, and implemented numerically using the projector in the SPGPE. This corrects the single particle cutoff for the compression of energy levels caused by the presence of the condensate. By finding consistent SPGPE parameters, we are able to set the value of the damping constant $\gamma$ through Eq.~(\ref{gamdef}), and thus determine all physical parameters of the theory. 

Comparing three c-field theories capable of describing vortex decay in different levels of approximation, we found that the SPGPE theory leads to the fastest decay in all cases considered. While this is expected as the SPGPE contains the other theories as special cases, in some cases the differences between theories are quite dramatic. At the lower temperature considered, complete decay of the vortex was not observed in the PGPE theory in either the 4:1 or 1:1 geometries. Vortex decay was observed at the higher temperature, providing another example of ergodicity emerging from purely Hamiltonian evolution. For all cases where a finite lifetime is observed, the DPGPE led to slower decay that the PGPE. This is caused by the rapid damping in DPGPE of thermally occupied classical degrees of freedom in the pure PGPE theory. In order for the damping to give sensible results, it is thus vital to include the associated stochastic element of the reservoir interaction, leading to the SPGPE theory. When the temperature is high enough to give a finite lifetime in PGPE theory, the calculated value is within a factor of two of the predictions of the SPGPE.

Applying the SPGPE theory to vortex decay over a range of temperatures we observe that the vortex lifetime is linearly dependent on $T/T_c$ at fixed $N_{\rm T}$, and varies by approximately an order of magnitude over the temperature range $0.78T_c\leq T\leq 0.93T_c$. The shortest lifetimes, calculated for $5\times 10^5$ $^{87}$Rb  atoms at $T/T_c=0.93$ in 4:1 and 1:1 geometries, are $\bar{t}=2$s and $\bar{t}$=1.5s respectively. At the high temperatures corresponding to these short lifetimes, the angular momentum gives a clear indication of diffusive decay of the vortex, evidenced by a long tail for $\bar{t}<t$.

We note that our results for the vortex lifetime also provide a lower bound for the lifetime of a persistent current of unit winding number, as any additional vortex pinning potential will further extend the vortex lifetime. Our results have been calculated using a theory that we expect to be a very good description of the high temperature Bose gas in \emph{quasi}-equilibrium, without any fitting parameters. We are thus optimistic that our predictions will be of quantitative value in comparing with future experimental measurements of the lifetimes of vortices and persistent currents. Future work will focus on the role of the scattering term in the full SPGPE~\cite{SGPEII,Bradley08a,Blakie08a}, and a more detailed treatment of the incoherent region dynamics.
\par
We are grateful to T. M. Wright for stimulating discussions regarding this work.
We acknowledge support from the \ASBS.
\appendix
\section{Hartree-Fock estimation of SPGPE parameters}\label{DOSdetails}
In this appendix we give details of our method for determining consistent control parameters for the SPGPE theory $\mu$, $\ecut$, as a function of temperature $T$, for a system constrained to have fixed total particle number $N_{\rm T}$. The method is based on a Hartree-Fock treatment of the trapped Bose gas. 

In the Thomas-Fermi regime, the condensate density is given by
\eqn{n_0(\x) = \left\{ \begin{array}{cc}  \left( \mu - V (\mbf{x})  \right)/U_0 & \quad \mu - V(\mbf{x}) \geq 0 \\
0 & \quad \mu - V (\mbf{x}) < 0 \end{array} \right\} ,}
where $\mu$ sets the condensate occupation $N_0=\int n_0(\x)d^3\x$.
Using Eq.~(\ref{HFH}) to describe thermal excitations, the Hartree-Fock density of states for the system
\eqn{\rho_{\rm{HF}}(\epsilon) = \int \frac{d^3 \mbf{x}d^3 \mbf{p} }{(2 \pi \hbar )^3} \delta \left(\epsilon - E_{\rm HF}(\x,\p)   \right), }
can be used to determine the total atom number for a given $\mu$. For the harmonic trap, this can be evaluated to give the analytical expression of the form
\eqn{\rho_{\rm{HF}} (\epsilon) = \frac{2}{\pi \hbar \bar{\omega}} \left[ 	I_- (\epsilon) + I_+ (\epsilon)	\right] ,} 
where $I_- (\epsilon)$ and $I_+ (\epsilon)$ are defined by
\eqn{I_- (\epsilon) &=&\Bigg\{\frac{u^3_- x}{4} - \frac{a_-u_-x}{8} \nonumber\\
&&- \frac{a^2_-}{8} {\rm{log}} (x + u_-)\Bigg\}\Big|^{x = \sqrt{2 \mu/ \hbar \bar{\omega}} } _{x = \sqrt{ {\rm{max}} \{ 0,a_- \} } } }
\eqn{ I_+(\epsilon) &=& \Bigg\{ - \frac{u^3_+ x}{4} + \frac{a_+ u_+ x}{8}\nonumber\\
&& + \frac{a^2_+}{8} {\rm{sin}}^{-1} \left( \frac{x}{\sqrt{a_+}} \right) \Bigg\}\Big|^{x = \sqrt{a_+}}_{x = \sqrt{2 \mu / \hbar \bar{\omega}} }}
and where $a_{\pm} = 2(\epsilon \pm \mu_{\rm{TF}} ) / \hbar \bar{\omega}$, and $u_{\pm} = \sqrt{a_{\pm} \mp x^2}$.

The total number of atoms in the system is given by
\eqn{\label{eqn:Ntotal} 
N_{\rm{T}} = \int_0^{\infty} d \epsilon\; \rho_{\rm{HF}} (\epsilon) n_{\rm{BE}} (\epsilon) + N_0 ,
}
where
\eqn{ \label{eqn:BE} 
n_{\rm{BE}}(\epsilon) = \frac{1}{e^{(\epsilon - \mu)/ k_{\rm{B}} T} - 1}.
}
By specifying the total number of atoms in the system \eeref{eqn:Ntotal} can be solved numerically to give $N_0$, and hence $\mu$, using the Thomas-Fermi chemical potential for the harmonic trap
\eqn{ \mu = \frac{\hbar \bar{\omega}}{2}  \left( \frac{15 N_0 a}{\bar{a}} \right)^{2/5} ,}
where $\bar{a} = \sqrt{ \hbar / m \bar{\omega} } $.  For temperatures above $T_c$, $\rho_{\rm{HF}}(\epsilon)$ becomes the standard single particle density of states. The SPGPE parameter $\mu(T)$ can then be found easily for temperatures above the transition using (\ref{eqn:Ntotal}), with $N_0 = 0$.

We also need to determine $\ecut$ so that the highest energy $\rC$-region modes satisfy our occupation condition $\ncut$.
However, some care is required to identify the value of $\ecut$ for use in the SPGPE.  The SPGPE simulations are most conveniently performed using a spectral numerical tecnique based on the harmonic oscillator eigenstates of the trap~\cite{Blakie08b}. Thus $\ecut$ is chosen as the highest oscillator eigenstate included in the $\rC$ region. The actual physical energy of the corresponding oscillator state in the SPGPE c-field is then shifted by interactions. The simplest way to connect $\ecut$ with $\epsilon_{\rm cutHF}$ is to require that the number of modes in the $\rC$-region is the same in either representation.

The Hatree-Fock density of states takes energies relative to the condensate, so from (\ref{eqn:BE}) with $\mu = 0$ we have
\eqn{\epsilon_{\rm{cutHF}}  = k_{\rm{B}} T {\rm{ln}} \left( 1 + \frac{1}{n_{\rm{cut}}} \right) }
valid for $T < T_{\rm{c}}$, where $\epsilon_{\rm{cutHF}} $ is the energy set by the HF density of states at mean occupation $\ncut$.
The relevant number of single particle states is given by integrating $\rho_{\rm{HF}}$ up to $\epsilon_{\rm{cutHF}} $ 
since the condensate compresses the low lying levels together without changing their number.
 $\ecut$ is then found by requiring that 
\eqn{\int^{\epsilon_{\rm{cutHF}}}_0 d\epsilon\; {\rho_{\rm{HF}}} (\epsilon)  = \int^{\ecut}_{\epsilon_0} d\epsilon\; \rho_{\rm{HO}} (\epsilon) ,}
where $\epsilon_0=\hbar(2\omega_r+\omega_z)/2$, and $\rho_{\rm{HO}}(\epsilon) = \epsilon^2/2(\hbar \bar{\omega})^3$.
For temperatures above the transition, the single particle density of states is applicable so $\ecut$ can be found directly from the Bose-Einstein distribution
\eqn{\epsilon_{\rm{cutHF}} \equiv  \ecut = k_{\rm{B}} T {\rm{ln}} \left( 1 + \frac{1}{\ncut} \right) + \mu.}

This set of equations can be rapidly evaluated to give $\mu(T),$ and $\ecut(T)$ [see Fig.~\ref{fig3}].

\bibliographystyle{apsrev4-1}
%

\end{document}